\documentclass[aps,pra,amsmath,amssymb,superscriptaddress,reprint]{revtex4-1}%,linenumbers

\usepackage{graphicx}% Include figure files
\usepackage{dcolumn}% Align table columns on decimal point
\usepackage{bm}% bold math
\usepackage[utf8]{inputenc}
\usepackage[T1]{fontenc}
\usepackage{booktabs, array, mathptmx, float, tabularx, booktabs, lipsum, amsmath, multirow}
\usepackage{siunitx, xcolor}
\usepackage{pifont}
\usepackage[version=4]{mhchem}
\usepackage[colorlinks,linkcolor=blue,anchorcolor=blue,citecolor=blue]{hyperref}
\usepackage[linesnumbered,ruled,vlined]{algorithm2e}
\begin{document}
% Use the \preprint command to place your local institutional report number in the upper righthand corner of the title page in preprint mode.
% Use the 'preprintnumbers' class option to override journal defaults to display numbers if necessary
%\preprint{}

%\title{Classical packet switching assisted entanglement distribution in quantum networks}
\title{Hybrid packet switching assisted by classical frame for entanglement-based quantum networks}
% \email, \thanks, \homepage, \altaffiliation all apply to the current author. Explanatory text should go in the []'s, actual e-mail address or url should go in the {}'s for \email and \homepage.

% \affiliation command applies to all authors since the last \affiliation command. The \affiliation command should follow the other information
\author{Hao Zhang}
\email[]{zhanghaoguoqing@163.com}
%\thanks{}
\affiliation{Purple Mountain Laboratories, Nanjing 211111, China}

\author{Yuan Li}
%\email{}
%\thanks{}
\affiliation{Purple Mountain Laboratories, Nanjing 211111, China}

\author{Chen Zhang}
%\email{zhangchen@pmlabs.com.cn}
%\thanks{}
\affiliation{Purple Mountain Laboratories, Nanjing 211111, China}

\author{Tao Huang}
%\email{htao@bupt.edu.cn}
%\thanks{}
%\altaffiliation{Department of Physics, Massachusetts Institute of Technology, Cambridge, Massachusetts, USA}
\affiliation{Purple Mountain Laboratories, Nanjing 211111, China}
\affiliation{State Key Laboratory of Networking and Switching Technology, Beijing University of Posts and Telecommunications, Beijing 100876, China}

% \author{Yunjie Liu}
% %\email{zpml4@163.com}
% %\homepage[]{Your web page}
% %\thanks{}
% \affiliation{Purple Mountain Laboratories, Nanjing 211111, China}
% \affiliation{State Key Laboratory of Networking and Switching Technology, Beijing University of Posts and Telecommunications, Beijing 100876, China}
% %Collaboration name if desired (requires use of superscriptaddress option in \documentclass). \noaffiliation is required (may also be used with the \author command).
% %\collaboration can be followed by \email, \homepage, \thanks as well.
% %\collaboration{}
% %\noaffiliation

\date{\today}

\begin{abstract}
One of the first problems of studying the quantum internet is how to realize quantum interconnection between users in a quantum network. To address above problem, by referencing the classical Internet, developing the packet switching of quantum networks is a promising way. In this paper, we propose a new hybrid packet switching for entanglement-based quantum networks assisted by classical frame. Different from the previous packet switching for quantum networks based on single photon, the frame used in our scheme is pure classical rather than the classical-quantum structure, and the transmission of classical and quantum signals over physical channels can be independent, which makes our scheme is also valid for quantum networks with heralded entanglement generation. Using our hybrid packet switching, the process of building entanglement channel between end nodes is analogous to the classical packet-switched networks, which provides an effective way to build large-scale packet-switched entanglement-based quantum internet. To verify the feasibility, we perform end-to-end entanglement distribution using our hybrid packet switching in a quantum network and simulate the fidelities of distributed state with respect to the number of hops. 
\end{abstract}

%\keywords{quantum}

\maketitle


\section{Introduction}\label{}
With the increase of quantum nodes, it is necessary to develope quantum networks to execute quantum information tasks, such as quantum communications \cite{BB84,E91,satelliteQKD1,QSDC1,QSDCTWO1,QSDCTWO2}, distributed quantum computing \cite{DQC1,DQC2,DQC3} and sensing \cite{Qsensing0,Qsensing1,Qsensing2,Qsensing3}. Different from the investigation of quantum communications on one quantum channel with two users, quantum networks have a large number of quantum channels and nodes. Therefore, the development of quantum networks faces some new problems and challenges arising from network itself \cite{QInternet1,QInternet2,QInternet3,QInternet4,QInternet5}. In a classical communication network, one of the first problems is how to realize the communication between users. To address above problem, the circuit switching is used in the early phase of classical networks \cite{CPS}. As users and tasks increase, the packet switching is proposed and developed rapidly to meet the utilization of networks \cite{CPS}, which leads to today's Internet. When we consider constructing large-scale quantum networks, it is so necessary to develope a quantum version of packet switching for running quantum networks as well as making quantum and today's classical networks coexist \cite{QPS}.

However, the case is so different and complex in the investigation of quantum networks due to different physical principles than classical networks, e.g. quantum superposition and entanglement have no classical counterpart. It is difficult to apply the techniques of classical networks to quantum networks directly \cite{QIPS,EAQN}. In recent years, in order to develope a packet switched quantum networks, classical-quantum hybrid frame is proposed with the classical header, trailer and quantum payload \cite{QPS}. By using this frame structure, the quantum signal can be transported from source to destination nodes successfully in a quantum network, which is analogous to classical packet switching. This framework is designed for quantum networks based on single photon. As an application, it can also be applied to distribute entanglement by transporting the entangled photons directly. However, in entanglement-based networks, some kinds of heralded entanglement generation used for constructing entanglement channels between adjacent nodes \cite{QR1,QR2,QR3}, such as mid-point Bell state measurement (BSM) \cite{QR1,QR2}, is not suitable for the previous packet switching based on hybrid frame. It is necessary to develope a more general packet switching for entanglement-based networks.

In this paper, we propose a new scheme of hybrid packet switching for entanglement-based quantum networks directly. By using our scheme, the entanglement channel can be extended from source node to destination node hop-by-hop assisting by classical frame, which is analogous to classical packet switching. Different from the previous packet switching for quantum networks \cite{QPS}, we make use of pure classical frame rather than classical-quantum hybrid frame, i.e. there is no quantum signal in the frame, and the transportation of classical and quantum signals over physical channels can be independent. This property makes our scheme valid for some cases beyond the previous packet switching in entanglement-based quantum networks, such as entanglement generation with mid-point BSM. Using our hybrid packet switching, the entanglement-based quantum networks can be executed in packet-switched model, and coexist with today's classical Internet well. To verify the feasibility, we perform the end-to-end entanglement distribution in a quantum network by using our hybrid packet switching technique and simulate the fidelities with respect to the number of hops. 

The article is organized as follows: In Sec.\ref{secCFAHPS}, we introduce entanglement-based quantum networks and classical packet switching in subsec. \ref{secEQN} and subsec. \ref{secCPS}, respectively. In Sec. \ref{secHPS}, we show the scheme of hybrid packet switching and introduce the requirements of hardware and collaboration. As an application, we perform the end-to-end entanglement distribution via our hybrid packet switching and simulate the fidelities in Sec. \ref{secEEED}. A summary is given in Sec. \ref{secsummary}.

\section{Quantum networks and classical packet switching}\label{secCFAHPS}

\subsection{Entanglement-based quantum networks}\label{secEQN}

\begin{figure}[]
    \includegraphics[width=\linewidth]{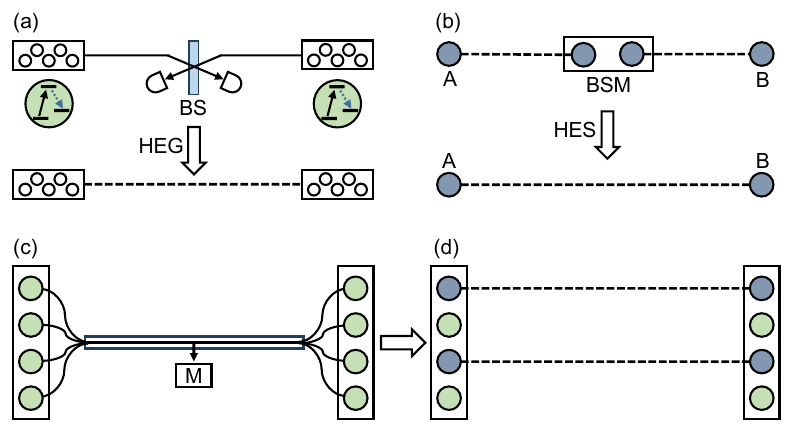}
    \caption{Quantum operations about entanglement in quantum networks. 
    (a) Heralded entanglement generation in two adjacent quantum nodes. The qubits are given by  atomic ensembles with three-level. 
    (b) Heralded entanglement swapping for extending entanglement channel. 
    (c) The physical channel of adjacent quantum nodes with multi-qubit connected by optical fibres. 
    (d) The construction of entanglement channels between the qubits of adjacent nodes. 
    BS, beam splitter; HEG, heralded entanglement generation; BSM, Bell states measurement; HES, heralded entanglement swapping; M, measurement. Dashed lines represent entanglement.}\label{QOQN}
\end{figure}

Quantum networks composed of many quantum nodes can be constructed for executing various quantum information tasks, such as quantum communications and distributed quantum computing. So far, quantum networks are divided into two types, i.e. based on single-photon and entanglement. Due to photon loss and operation errors, entanglement can not be distributed between long-distance quantum nodes directly. Therefore, to solve above problem, quantum repeaters are created by using heralded entanglement generation and entanglement purification or quantum error correction to overcome photon loss and operation errors, respectively, and performing entanglement swapping to extend entanglement distance \cite{QR0,QR1,QR2,QR3,QR4,QR5}. When many quantum nodes are connected into a quantum network, quantum devices for routing quantum signal or entanglement is required. Here, we consider entanglement-based quantum networks and focus on the class of quantum repeaters based on heralded entanglement generation and entanglement swapping \cite{QR2,QR3}. Before executing protocols of application layer, one should construct an end-to-end entanglement channel between users in entanglement-based quantum networks \cite{ERouting0,ERouting1,ERouting2,ERouting3,ERouting4,ustcQN}. In entanglement-based quantum networks, some relevant concepts are introduced as follows.

\emph{Heralded entanglement generation}. In quantum networks, it is a process of constructing entanglement channels between qubits in adjacent nodes \cite{QR1,QR2,QR3}. Different from transporting half photon of a Bell state from one node to another, heralded entanglement generation puts photons from two nodes together to perform a BSM and produces a desired entangled state.  As shown in Fig. \ref{QOQN} (a), the example is a mid-point scheme in which the end nodes are two atomic ensembles with three-level system. The photons are produced by stimulated Raman process and transported to a beam splitter. The success of entanglement generation is heralded by trigger of one detecter. By BSM and postprocessing, the two atomic ensembles can be prepared with a desired Bell state. 

\emph{Entanglement swapping}. It is a process of transferring entanglement to other nodes. Actually, it is a quantum teleportation on entanglement \cite{teleportation}. An example is shown in Fig. \ref{QOQN} (b), after performing a BSM between two qubits in the middle node, the entanglement is produced on residual qubits. In quantum networks, this operation is used for extending entanglement channels.

\emph{Entanglement purification}. It aims at improving the fidelity of nonlocal entanglement by local operations and classical communications \cite{BBPSSW,husheng,ursinprl,vqcep,shengscpma}. This process should consume additional nonlocal entanglement and is usually used for mitigating operation errors in quantum communications. A typical example of entanglement purification protocol is to perform bilateral controlled-NOT gates and parity check on two-copy mixed states in both quantum nodes \cite{BBPSSW,husheng}.

\emph{Quantum error correction}. It can recover the state suffered from errors by redundant encoding \cite{qec1,qec2,zengqec}. By measuring the ancilla qubits and extracting the syndrome information of data qubits, the state can be recovered by corresponding error correcting operations. It is proposed for fault-tolerant quantum computation and also extended for supporting long-distance quantum communications.

In the future practical quantum networks, there are sufficient quantum memories in a quantum node to satisfy communication requirements \cite{QM1,QM2,QM3}. As shown in Fig. \ref{QOQN} (c) and (d), more than one entanglement channels between adjacent nodes are constructed via heralded entanglement generation.

%!!!!!!!!!!!!!!all the figures should be translated to pdf in the last version!!!!!!!!!!!!!!!! 
\begin{figure}[]
    \includegraphics[width=\linewidth]{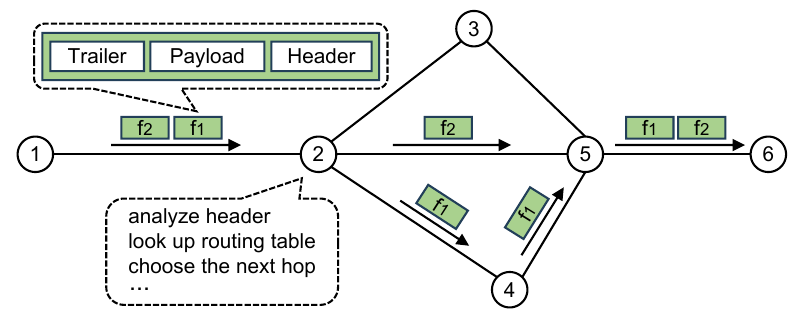}
    \caption{Classical packet switching in the Internet. The classical frame composed of header, payload and trailer is used for transmitting messages from sender node to receiver node. Routers have function of routing classical frame to the next hop according to the information of header. Rectangle labelled with $\text{f}_1$/$\text{f}_2$ is classical frame.}\label{CPS}
\end{figure}

\subsection{Classical packet switching and frame}\label{secCPS}
Packet switching is a well known approach widely used for transmitting messages in today's Internet. Different from circuit switching, the message in sender node is divided into many parts which will be transmitted to the receiver node independently in a network in packet switching. During transmission, the classical frame plays a crucial role for promising that each part of message can arrive at right destination. The packet switching and classical frame of today's Internet are shown in Fig. \ref{CPS}. The rectangles labelled with $\text{f}_1$ and $\text{f}_2$ are classical frames. When frames are sent to the network, they will be transmitted independently. Therefore, they may be routed to different hops but get together at end node. Due to covering different paths in a network, the arrival order of frames are inconsistent with departure order, e.g. $\text{f}_2$ arrives at end node 6 before $\text{f}_1$. The classical frame is composed of header, payload and trailer. Briefly, the header is encoded with information for routing, i.e. address of sender and receiver, error correction and etc. With those information, the frame can be routed form sender to receiver node hop by hop successfully. The payload is each part of message which needs to be sent, and the trailer is a signal for the end of the frame. The detailed protocol is that when a frame arrives at a router, it's header will be analyzed. Based on the information read from the header, the router should look up its local routing table and decide the next hop which the frame will be sent. In practice, the frame transmitted in a network needs a complex network protocol stack composed of protocols of each layer.

\section{The approach of hybrid packet switching for quantum networks}\label{secHPS}

\subsection{Constructing entanglement channels between adjacent nodes}\label{seccase12}

\subsubsection{Case I: non-oriented entanglement generation}\label{seccase1}
\begin{figure}[]
    \includegraphics[width=\linewidth]{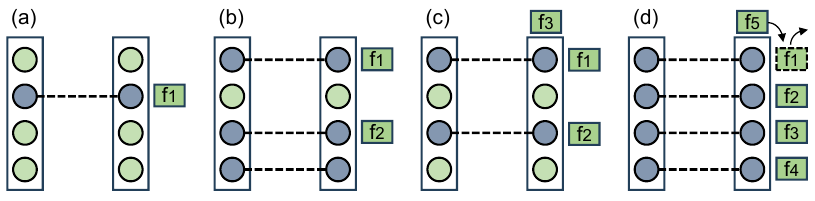}
    \caption{The four cases of allocating entanglement channels between adjacent nodes for the arriving frames in non-oriented entanglement generation. 
    (a) The number of frames equals to entanglement channels. 
    (b) The number of frames is less than entanglement channels. 
    (c) The number of frames is more than entanglement channels. 
    (d) The number of frames is more than memory qubits. 
    The $\text{f}_{i}$ is the \emph{i}-th frame arriving at this node.}\label{case1}
\end{figure}

There are two different cases of constructing entanglement channels between adjacent nodes. The first one is non-oriented entanglement generations, i.e. the processes of entanglement generation and handling classical frame are independent. For simplicity, we assume that heralded entanglement generations are always executed between all the adjacent nodes to keep enough entanglement channels to be used. When memory qubits are prepared with entangled state, the entanglement generation between those qubits is suspended until the qubits are released. There are two cases for releasing qubits, one is accomplishing the task, such as BSM and entanglement purification. The other case is the fidelity of entanglement is lower than a threshold and entanglement should be discarded and reproduced again. As shown in Fig. \ref{case1} by considering four quantum memories in each quantum node. Different cases of allocation are discussed in Fig. \ref{case1} (a)-(d). In the case (a) where the number of frames is the same with entanglement channels, we give the instance in which a classical frame $\text{f}_{1}$ arrives at right node from left and only one entanglement channel is constructed successfully now. The channel should be allocated for $\text{f}_{1}$ directly. The case (b) whose frames are less than entanglement channels is shown with three prepared entanglement channels and two arriving classical frames. The frame will be allocated with two entanglement channels in order of arriving time. In the case (c), the arriving frames are more than entanglement channels and the quantum memories also have free qubits, i.e. there are two entanglement channels but with three classical frames $\text{f}_{1}$, $\text{f}_{2}$ and $\text{f}_{3}$ in Fig. \ref{case1} (c). According to the arriving time, $\text{f}_{1}$ and $\text{f}_{2}$ are allocated with those entanglement channels, but $\text{f}_{3}$ should queue for another entanglement channel to be constructed. The last case (d) is that the frames are more than memory qubits. In Fig. \ref{case1} (d), we show a new classical frame $\text{f}_{5}$ arrives at the node but all four entanglement channels are occupied. It is a little complex in this case. At this point, the frame $\text{f}_{5}$ should queue. After the one of four frames leaves and arrives at next node, the BSM in this node should be performed. Now the memory qubit can be released and new entanglement channel is constructed again for $\text{f}_{5}$. This process may consume much time. If the queueing time is so long that exceed the time threshold, the frame should be discarded. This problem can be mitigated by adding enough quantum memories and optimizing the traffic flow of frame in classical networks. It's worth noting that when the classical frame leaves, the quantum memories in this node are occupied. After executing BSM to extend entanglement channel to next node successfully, the occupied quantum memories are released and subsequently executed with heralded entanglement generation.

\subsubsection{Case II: classical frame oriented entanglement generation}\label{seccase2}

\begin{figure}[]
    \includegraphics[width=\linewidth]{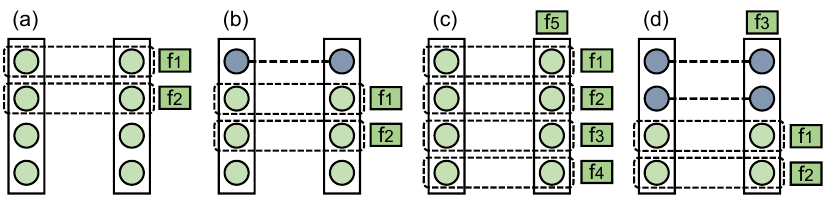}
    \caption{The four cases of allocating quantum memories between adjacent nodes for the arriving frames in classical frame oriented entanglement generation. 
    (a) All the quantum memories are free and the number of frames is less than memory qubits. 
    (b) Some quantum memories are occupied and the number of frames is less than free memory qubits. 
    (c) All the quantum memories are free and the number of frames is more than memory qubits. 
    (d) Some quantum memories are occupied and the number of frames is more than free memory qubits. 
    The $\text{f}_{i}$ is the \emph{i}-th frame arriving at this node.}\label{case2}
\end{figure}

The second case we discussed here is classical frame oriented entanglement generations, i.e. generating entanglement according to the frame. The protocol is that when the frame arrive at a node, the free quantum memories in current node and the previous adjacent node will be allocated to this frame for constructing entanglement channels. Before the frame arrive at the node, there is no free prepared entanglement channels between those two adjacent nodes. Compared with non-oriented case which is assumed heralded entanglement generations are always executed between all the adjacent nodes, this case should spend more time on building entanglement channels, but less resource will be costed in a quantum network with low number of request. In Fig. \ref{case2}, we discuss four cases of allocating quantum memories for the arriving frames. In the case (a), all the quantum memories are free, and the number of arriving  frames is less than quantum memories. The frame will be allocated with a pair of quantum memories between adjacent nodes to build entanglement channels according to the order of arriving. In the case (b), some quantum memories are occupied with entanglement channels for other request, and residual free quantum memories are more than arriving frames. The frame will be allocated with quantum memories except for the occupied ones. The case (c) is all quantum memories are free and the arriving frames are more than memory qubits. The last frame $\text{f}_{5}$ should wait for the free quantum memories. The last case is in Fig. \ref{case2} (d), the partial quantum memories are busy, and the free memory qubits are less than the arriving frames. This case is similar to the case (c), the frame $\text{f}_{3}$ should queue for a pair of new free memory qubits to be allocated.

\subsection{Extending the quantum channel of classical frame}

\begin{figure}[]
    \includegraphics[width=\linewidth]{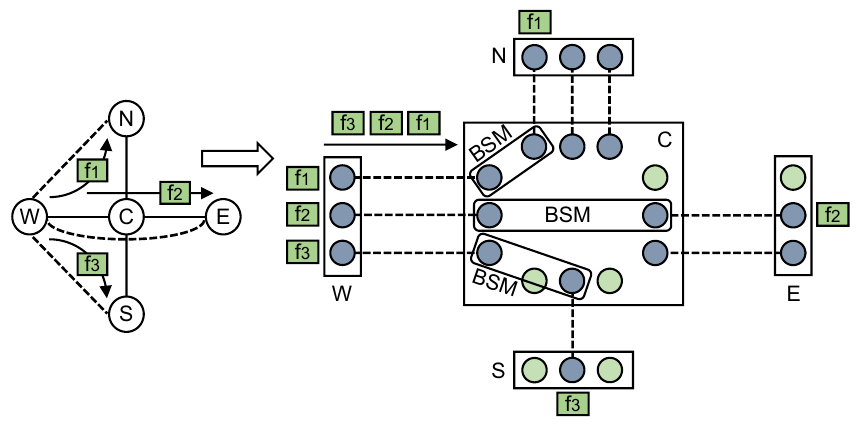}
    \caption{The extension of entanglement channels in quantum switch via entanglement swapping. BSM is Bell state measurement. Node W, N, E, S, and C are the western, northern, eastern, southern and central nodes, respectively.}\label{extension}
\end{figure}

The above two cases can share the same approach for extending quantum channels via entanglement swapping. After allocating quantum channels to the arrived classical frame by using the protocol introduced in sec. \ref{seccase12}, the previous node traveled by the frame should perform a BSM on entangled qubits allocated to the corresponding same frame and send the measurement outcome to the end node to determine the final entangled state. The swapped entanglement also can be corrected hop-by-hop based on BSM outcome, but performing correction on the final end node is more convenient. To show the approach clearly, we show a two-hop cross structure composed of five nodes in Fig. \ref{extension}. We consider the case that three classical frames, i.e. $\text{f}_{1}$, $\text{f}_{2}$ and $\text{f}_{3}$, come from the western node to the central node and subsequently are routed to the northern, eastern and southern nodes, respectively. When the frame arrives the central node, a new entanglement channel is allocated. Subsequently, the frame will be routed to the second node by classical router, e.g. the frame $\text{f}_{1}$ travels to the northern node, a new entanglement channel between central node and next node is allocated, and the central node will perform a BSM on qubits belonging to two entanglement channels allocated to this frame, i.e. the BSM in the upper left corner of central node in Fig. \ref{extension}. Similarly, the quantum channels of frame $\text{f}_{2}$ and $\text{f}_{3}$ are extended in this way. In a linear repeater chain, the above process is only the extension of quantum channel. However, in a network, the quantum node with multi-path can play a role in quantum router for routing entanglement. In Fig. \ref{extension}, three entanglement channels will be built between node W and N, W and E, and W and S for the frame $\text{f}_{1}$, $\text{f}_{2}$ and $\text{f}_{3}$, respectively. Extending the system to a practical network, entanglement can be routed from source node to destination node hop-by-hop under the guidance of the routed classical frame.
However, in practice, the entanglement is usually not one pair, and probably multi-pair entangled states between two blocks of memory qubits. Because, firstly, some kinds of tasks require more than one pair of entangled state, and secondly redundant entanglement should be used for correcting errors, such as entanglement purification and quantum error correction. Therefore, the extension of entanglement channels in Fig. \ref{extension} is executed between two blocks of memory qubit in multi-pair case.

BSM is crucial in entanglement swapping and can have different ways to be executed according to the topology of the network. If the topology and the distance between all nodes of network are fixed and known clearly in advance for all nodes, the times of transporting photon in all physical channels can be estimated. In this case, when the frame leaves the node, the BSM can be executed after a fixed time delay without the order from the next node for saving time. This scheme also requires that the node has ability of calculating which entanglement channel is allocated for the frame when the frame arrives the next node according to the information about the quantum resource between adjacent nodes. If the topology is dynamic with randomness, the performance of BSM can be started by accepting the order from the next node.

\subsection{Hardware and collaboration}

\begin{figure}[]
    \includegraphics[width=\linewidth]{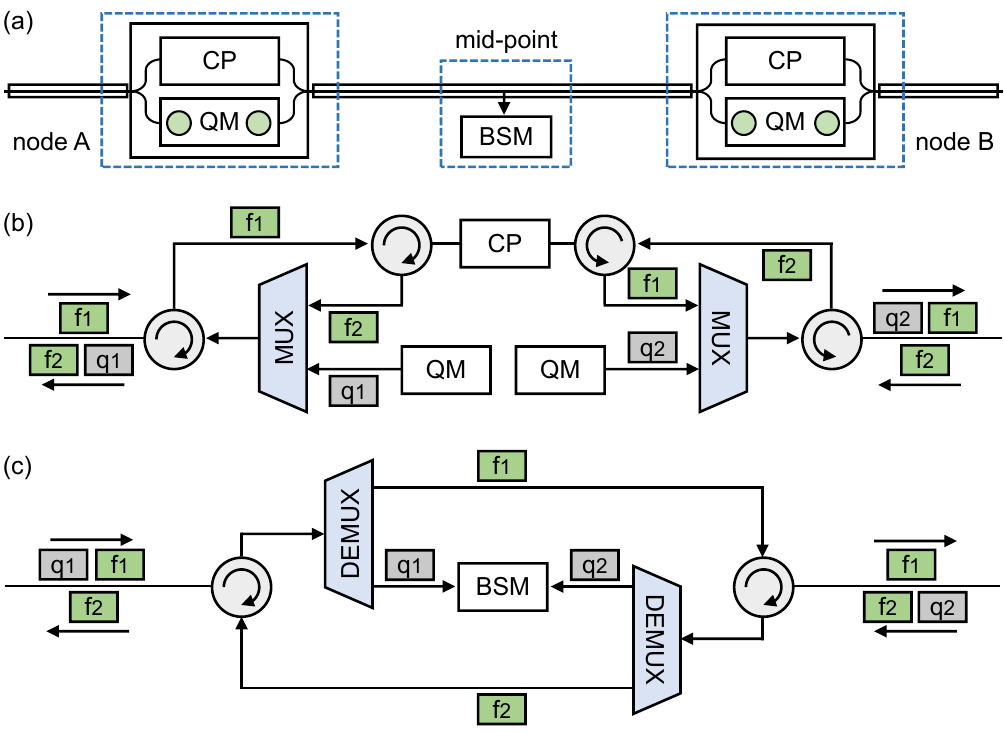}
    \caption{A feasible hardware proposal for processing classical frame and mid-point heralded entanglement generations in quantum networks. 
    (a) The schematic diagram of two adjacent nodes with entanglement generation via mid-point BSM. 
    (b) The hardware structure of the node A and B. 
    (c) The hardware structure of the mid-point BSM. 
    MUX, multiplexer; DEMUX, de-multiplexer; BSM, Bell states measurement; CP, classical processor; QM, quantum memory. The circle with a arrow inside is circulator.}\label{hardware}
\end{figure}

The hybrid packet switching contains both classical and quantum signals. For simplicity, the classical and quantum signals can be processed by two independent physical channels in principle. Considering coexistence with current classical networks, the classical and quantum signals also can share the same physical channel. Here, we introduce a feasible hardware proposal for above purpose in optical fibre based networks. The wavelength-division multiplexing (WDM) can be used for classical and quantum signals, and the different signals in classical or quantum domains can be operated using the time-division multiplexing (TDM). For example of heralded entanglement generation via mid-point BSM shown in in Fig. \ref{hardware} (a), feasible hardware proposals for nodes and BSM are shown in Fig \ref{hardware} (b) and (c), respectively. In each node, the transportation of classical signal is two-way. But the quantum signal is emitted from quantum memory and travels to the mid-point. To realize above function, two multiplexers (MUX)  and four circulators are used. An instance is shown in Fig \ref{hardware} (b) with two classical frames $\text{f}_{1}$, $\text{f}_{2}$ and quantum signals $\text{q}_{1}$, $\text{q}_{2}$. The classical frame $\text{f}_{1}$ is routed from left to upper classical processor by circulator and multiplexed to the right fibre by a MUX. The process is the same for the frame $\text{f}_{2}$ from right. In Fig \ref{hardware} (c), the hardware for the mid-point BSM is proposed. The left input $\text{f}_{1}$ and $\text{q}_{1}$ are separated by a DEMUX. Subsequently, the classical signal travels to the right directly and couples to the fibre by a circulator, and the quantum signal transfers to the instrument of BSM. The right signals are designed with lower paths.

Besides hardware, the collaboration between classical and quantum processors is also significant. In the case of classical frame oriented entanglement generation, when the classical frame arrive the node, classical processor will analyze its information and send orders to quantum processors to start heralded entanglement generation. When the mid-point completes the measurement, the outcome should send to bilateral node to inform classical and quantum processors. The similar process is required in the extension of quantum channel. Also in the management of quantum memory, the classical system should reasonably allocate memory qubits to classical frame. The quality of the collaboration will determine the speed of entanglement distribution and further effect the efficiency of whole quantum networks.

\section{End-to-end entanglement distribution}\label{secEEED}

\subsection{The protocol of entanglement distribution}
\begin{figure}[]
    \includegraphics[width=8.0cm]{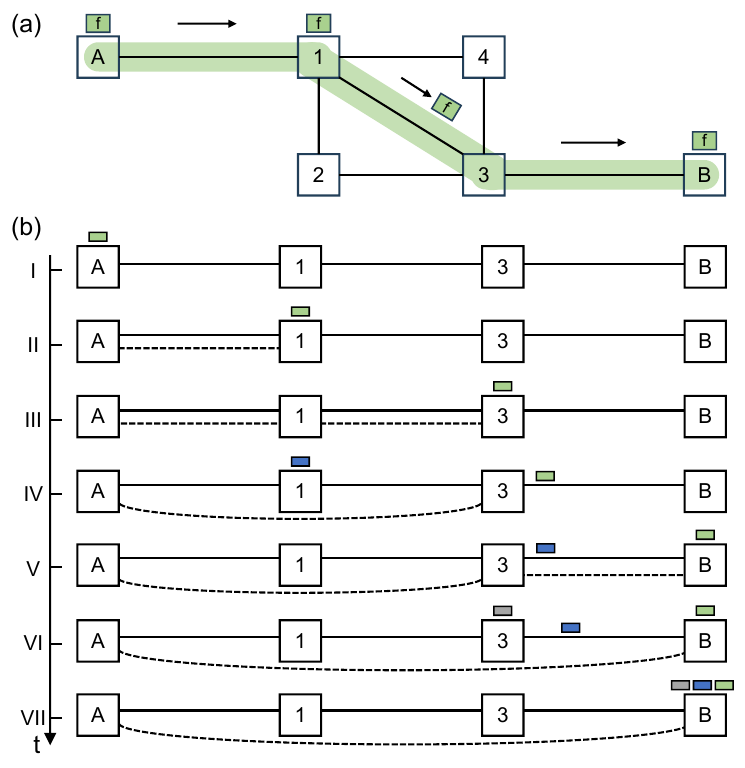}
    \caption{End-to-end entanglement distribution with single request between end node A and B in a quantum network. 
    (a) The topology of a simple quantum network.
    (b) The detailed processes of end-to-end entanglement distribution via hybrid packet switching over the path, i.e. A-1-3-B shown in (a), chosen by routers. The rectangle labelled with f stands for a classical frame. The solid line is physical channel where both classical and quantum signals transport, such as optical fibre, and dashed line represents allocated entanglement channel.
    }\label{endED}
\end{figure}

To perform quantum information tasks in a quantum entanglement network, the first phase is to realize end-to-end entanglement distribution between arbitrary users. In this section, we perform the construction of end-to-end entanglement channels by using our hybrid packet switching. For simplify, we consider the non-oriented case of entanglement generation in sec. \ref{seccase1} and single request between Alice and Bob shown in Fig. \ref{endED}. The detailed processes in time axis of Fig. \ref{endED} (b) and corresponding operations are given with both classical and quantum parts as follows.

\begin{itemize}
    \item[I,] When a request of entanglement distribution between node A and B is executed, node A produces a classical frame with information including node B's address and other uses. The frame will be sent to the physical channel which connects node 1. For quantum part, all the memory qubits are always executing heralded entanglement generations between adjacent nodes independently. If the memory qubits are entangled successfully, the entanglement generation is suspended until the qubits are released.
    \item[II,] After traveling over physical channel, the classical frame arrives at node 1 and the header is analyzed by classical processor. By looking up the local routing tables, the next node is chosen with node 3. Subsequently, the frame will be sent out. In quantum part, when the classical processor processes the arriving classical frame, it allocates a prepared quantum channel between node A and 1 for the frame. Now, the corresponding quantum memories are occupied and can not be allocated for other frames. The dashed line between node A and 1 in Fig. \ref{endED} (b) is allocated entanglement channel.
    \item[III,] The classical frame arrives at node 3. The header is analyzed by classical processor and the next hop is chosen with the end node B. One of quantum channels between node 1 and 3 is allocated for the frame. The entanglement swapping will be started in node 1. Subsequently, the classical frame will be sent out to the next node.
    \item[IV,] The classical processor in node 1 sends order of starting entanglement swapping to quantum part. BSM is performed on memory qubits allocated to the corresponding frame and the measurement outcome is sent via classical message (blue rectangle) to end node B. After the BSM, the quantum channel of the frame is extended between node A and 3 (the dashed curve connecting node A and 3) and the corresponding memory qubits in node 1 are released for building new entanglement channels. Now, the classical frame is traveling over the physical channel between node 3 and B with the case that the time of BSM is less than the time of transmitting classical frame over the next physical channel.
    \item[V,] The classical frame arrives at end node B. The header is processed. A quantum channel between node 3 and B is allocated for the frame.
    \item[VI,] The classical processor sends order to quantum processor to start entanglement swapping in node 3. Subsequently, BSM is performed on memory qubits allocated to the corresponding frame in node 3 and the outcome is sent to node B via classical communication (gray rectangle). After accomplishing BSM, the end-to-end entanglement channel is built between node A and B. 
    \item[VII,] The last classical message carrying with BSM outcome arrives at end node B. According to all the BSM outcomes from the middle nodes, the correction on entangled qubit is executed and end-to-end entanglement distribution is completed between node A and B. A feedback message of completing the task is sent back to node A.
\end{itemize}

In above processes, we just consider the case of allocating one entanglement channel (only one pair of entangled state) for the classical frame. In practice, more than one entanglement channels should be allocated for entanglement purifications or quantum error corrections. Because operation errors and decoherence of quantum memories will decrease the fidelity of entanglement. If the fidelity is not satisfy requirements, the entanglement distribution should be executed again. The distribution of multi-pair entanglement also can be used for protect entanglement channel from the failure of BSM. In above example of entanglement distribution, we only consider the case whose all BSM are successful. If only one pair of entanglement is distributed and the BSM in a node fails, the distribution should be performed again.

In quantum networks with many users, multi-request of end-to-end entanglement distribution between multiple users is a common case. Our scheme allows to execute multiple requests simultaneously, which is analogous to classical Internet based on packet switching. When more than one frames arrive at a node from the same physical channel, the entanglement channels will be allocated in order of arriving time of the frame according to the protocol in Fig. \ref{case1}. If frames are more than quantum memories, the queuing mechanism will be used. If two frames from different nodes arrive at a central node and will be routed to different next nodes, there is no influence on each other on allocating quantum resource and only the processing of classical frame should be optimized in classical processors. In our scheme, besides user's end node and the node which the frame arrives at, the quantum memories of the last node which the frame left from should be occupied for performing entanglement swapping. As shown in Fig. \ref{endED} (a), when the frame arrives at node B, the quantum memories corresponding to entanglement channels belonging to the frame in node 1 are free, but for node 3, the quantum memories are busy until accomplishing entanglement swapping.

\subsection{The simulations of fidelities}\label{secfidelity}

To show the effect from the increasing number of hop, we simulate the fidelities of end-to-end entanglement distribution with the case of non-oriented entanglement generations. The fidelity used in our simulations is defined by 
\begin{eqnarray}\label{fid}
    F(\rho_{0},\rho_{1})=\left(\text{Tr}\sqrt{\sqrt{\rho_{1}}\rho_{0}\sqrt{\rho_{1}}}\right)^{2}.
\end{eqnarray}
Here, $\rho_{0}$ is the density matrix of target state and $\rho_{1}$ is the practical density matrix of the system evolving with noises.
The target state distributed between two end nodes is Bell state 
\begin{eqnarray}\label{Bell}
    |\Psi_{+}\rangle = \frac{1}{\sqrt{2}}(|00\rangle + |11\rangle).
\end{eqnarray}
In our simulations, the final fidelity of distributed entangled state is the state after entanglement swapping, i.e. the state of the time when the last classical message carrying with outcomes of BSM arrives at end node, rather than completing last BSM. Because before the outcomes of BSM arrive at end node, the entangled state can not be determined.

\begin{figure}[]
    \centering
    \includegraphics[width=\linewidth]{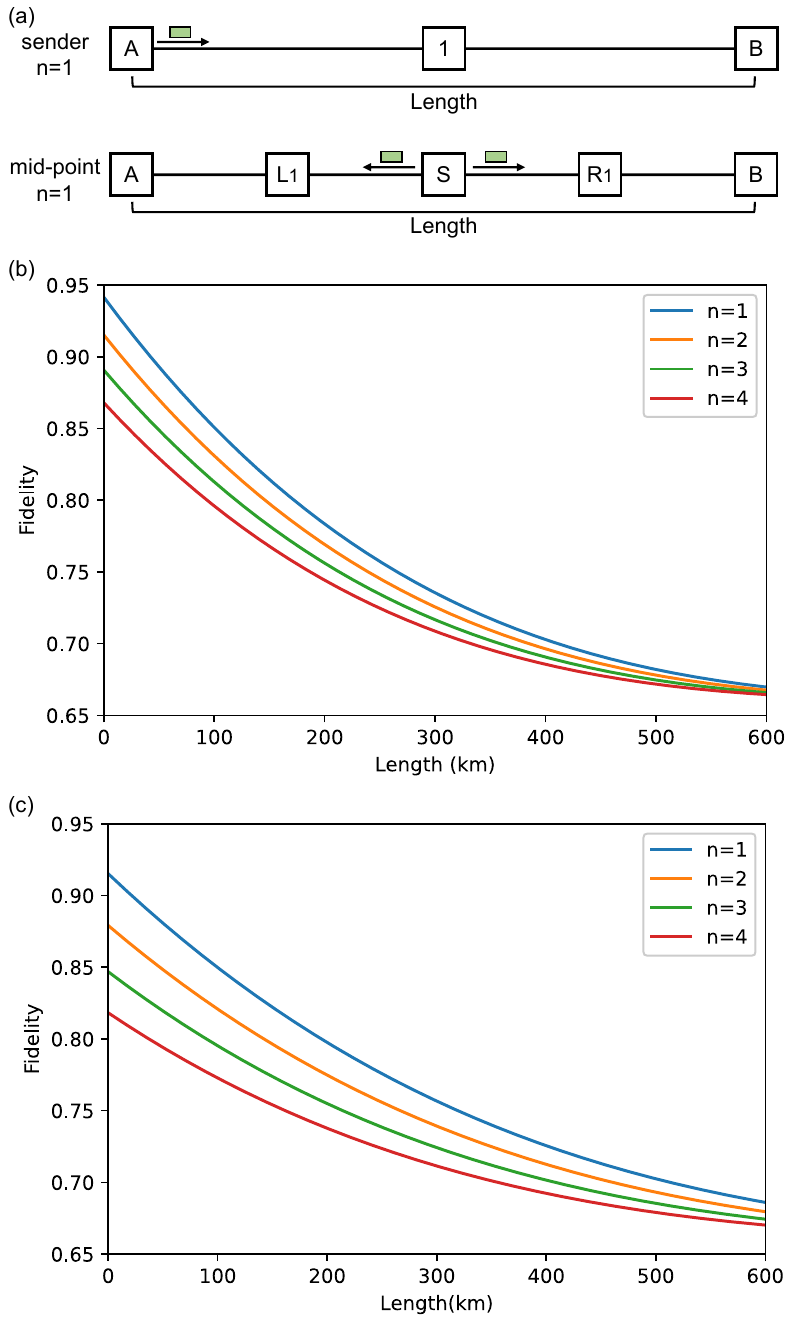}
    \caption{The simulations of fidelities of end-to-end entanglement distribution by using our scheme. 
    (a) The two cases of entanglement distribution, i.e. distribution from sender and mid-point. 
    (b) The fidelities of distribution from sender with respective to quantum nodes and length. 
    (c) The fidelities of distribution from mid-point with respective to quantum nodes and length.} \label{fidelity}
\end{figure}

The processing time of classical frame and allocating entanglement channel in each node are labelled with $t_{\text{p}}$ and $t_{\text{a}}$, respectively. The time for BSM is $t_{\text{bsm}}$. In our simulations, we assume that the processing time of classical frame is $t_{\text{p}} = 125\mu $s \cite{QPS}, and $t_{\text{a}} = t_{\text{bsm}} = t_{\text{p}}$ for simplicity. All the classical messages carrying the outcome of BSM have no fixed path in its transmission in classical packet-switched networks in practice, but they are also assumed that following the classical frame of request in simulations. The velocity of signal in optical fibre is chosen with $c_{\text{f}} \simeq 0.21$km/$\mu$s. All the quantum memories in quantum nodes are suffered from the $\text{T}_{1}\text{T}_{2}$ noise model, where $T_{1}$ and $T_{2}$ are the relaxation and dephasing time of memory qubit, respectively \cite{{Netsquid}}. Here, two memory parameters are chosen with $T_{1} = T_{2} = 3$ms. To simulate the effect from the increasing number of hop in physical channel, we further assume that the allocated entanglement channels between adjacent nodes are prepared with $|\Psi_{+}\rangle$ of $100\%$ fidelity and all the outcomes of BSM in entanglement swapping are the case of $|\Psi_{+}\rangle$. Two cases of entanglement distribution, i.e. distribution from sender and mid-point \cite{QPS}, are considered and results are given in Fig. \ref{fidelity}. The schematic diagrams of case $n=1$ for distribution from sender and mid-point are shown in Fig. \ref{fidelity} (a). $n$ in the figure is the number of node in physical channel between start point and end point. For instance, $n=1$ indicates there is one node between sender and receiver nodes in the case of distribution from sender. But in the case of distribution from mid-point, there are two nodes, which are on the left and right sides of start point, between sender and receiver. The results in Fig. \ref{fidelity} (b) and (c) indicate that as the length increases, the fidelities decrease. The reason is that it should taken much time in transmission of signal over longer length physical channel.
And another same results in Fig. \ref{fidelity} (b) and (c) is that the increasing number of hop $n$ reduces the fidelities. Because in our simulations, entanglement purification or quantum error correction is not performed to recover the fidelity in each node. As the number of hop increases, the total time consumed on processing classical frame $t_{\text{p}}$ and performing BSM $t_{\text{bsm}}$ becomes longer. Longer time indicates the entangled state stored in quantum memories should suffered from much decoherence and has lower fidelity. Therefore, in practical long-distance quantum networks, entanglement purification or quantum error correction is a promising way to protect the distributed state from noise. Comparing results of $2n$ in Fig. \ref{fidelity} (b) and $n$ in (c), we find that the case of distribution from mid-point is better than distribution from sender as the length increases. Because compared with distribution from sender, it saves half time in transmitting signals in distribution from mid-point.

%\textbf{if $\rho_{0}$ is a pure state, the fidelity is $F(\rho_{0},\rho_{1})=\text{Tr}(\rho_{0}\rho_{1})$}. 

% \section{Discussion}\label{secdiscussion}

% compared with three works from MIT, and a work from ustc. Ours is packet switching based. 1. compatible to classical network. 2. capacity is good because of packet switching, quantum memories are released after entanglement swapping.

% compared to packet switching with hybrid frame, 1, we need quantum memories, 2, suitable for entanglement swapping network, previous model is complex in entanglement swapping, 3, no hybrid architecture (no need of hardware for producing hybrid frame), 4, classical and quantum signals are transported and operated independently in space, just has some relationship in time. 5, good scalability due to independent, the quantum signal is not limited for CV???. 6 also can transport classical massages, but the previous one can not.  Another way for building entanglement-based quantum Internet with packet switched network.  

% is little different with conventional packet switching, because the quantum channel between adjacent nodes should be occupied for some time (waiting for next entanglement channel, i.e. transporting and processing classical frame, and entanglement swapping,) 

% can be applied to the entanglement distribution \cite{ustcQN}

\section{Summary}\label{secsummary}
In conclusion, we have proposed a new hybrid packet switching for entanglement-based quantum networks. Assisted by classical frame, end-to-end entanglement distribution can be accomplished by extending and routing entanglement channel from sender to receiver hop by hop. Different from previous packet switching based on classical-quantum frame, our scheme makes use of pure classical frame and transmission of classical and quantum signals can be independent, which makes it is also valid for quantum networks with heralded entanglement generation. Our work provides a new way for building large-scale packet-switched entanglement-based quantum internet.

\begin{acknowledgments}
We thank Stephen DiAdamo, Yu-Ming Xiao and Xuan Mao for helpful discussion. %This work is supported by xxx. 
\end{acknowledgments}

% \appendix
% \section{Quantum teleportation and entanglement swapping}\label{apppsbz}
% Entanglement swapping based on BSM actually is the quantum teleportation of a entangled state. A composite system with two $|\Phi^{+}\rangle$ is described by 
% \begin{eqnarray}\label{}\nonumber
%     |\psi\rangle&=&|\Psi^{+}_{12}\rangle|\Psi^{+}_{34}\rangle\\\nonumber
%     &=&\frac{1}{2}[|\Psi^{+}_{14}\rangle|\Psi^{+}_{23}\rangle+|\Psi^{-}_{14}\rangle|\Psi^{-}_{23}\rangle\\\nonumber
%     &&+|\Phi^{+}_{14}\rangle|\Phi^{+}_{23}\rangle+|\Phi^{-}_{14}\rangle|\Phi^{-}_{23}\rangle].
% \end{eqnarray}
% Therefore, a BSM on qubit 2 and 3 will project the subsystem composed of qubit 1 and 4 into one of four Bell states according to the outcome of BSM. For instance, the outcome of BSM is $|\Psi^{+}_{23}\rangle$, the state of qubit 1 and 4 is $|\Psi^{+}_{14}\rangle$. If the outcome is other Bell state, one should apply a local single qubit operation on qubit 1 or 4 to obtain the state $|\Psi^{+}_{14}\rangle$.

%\bibliographystyle{prl}
%\bibliography{references}

\end{document}